\documentclass[a4paper,conference]{IEEEtran}
\IEEEoverridecommandlockouts
\usepackage{cite}
\usepackage{amsmath,amssymb,amsfonts}
\usepackage{graphicx}
\usepackage{xcolor}
\usepackage{booktabs}
\usepackage{url}
\usepackage{algorithmic}
\usepackage{subcaption}
\renewcommand\IEEEkeywordsname{Keywords}

\renewcommand{\baselinestretch}{0.998}

\begin{document}

\title{Leveraging UAV Autonomy for Minimum 4D Flight Authorization Volumes \\\vspace{10pt}}

\author{Christian Vitale, Yiannis Grigoriou, Panayiotis Kolios, and Georgios Ellinas 
\thanks{C. Vitale, Y. Grigoriou, and G. Ellinas are with the Department of Electrical and Computer Engineering and the KIOS Research and Innovation Center of Excellence (KIOS CoE), University of Cyprus, Nicosia, Cyprus. P. Kolios is with the Department of Computer Science and the KIOS CoE, University of Cyprus. Emails: {\tt\small \{vitale.christian, grigoriou.yiannis, pkolios, gellinas\}@ucy.ac.cy}%
\newline
This work was supported by the Border Management and Visa Policy Instrument (BMVI), co-financed by the European Union and the Republic of Cyprus (BMVI/2021-2022/SA/1.2.1/015) (project REACTION). It was also partially supported by the European Union's Horizon Europe research and innovation program under grant agreement No 101187121 (EUSOME) and by the Republic of Cyprus through the Deputy Ministry of Research, Innovation and Digital Policy.}
}

\maketitle

\begin{abstract}
The increasing UAV traffic in urban areas has prompted the creation of U-space, an EASA framework for safe and efficient unmanned aerial vehicle (UAV) operations. Within this context, this work presents a flight authorization framework that leverages autonomous UAVs, using their motion models and control characteristics to improve authorization efficiency. In the proposed framework, probabilistic spatial-temporal envelopes are generated to predict future UAV locations within a desired confidence level, and this information is then used to determine the minimum 4D operational volumes that form a valid authorization request for the mission. By reserving only the necessary airspace, the approach enhances capacity and supports simultaneous UAV operations. Simulations comparing the proposed method with a conventional rule-based strategy demonstrate consistently more compact and efficient airspace reservations across a range of mission types.
\end{abstract}


\section{Introduction}
The usage of unmanned aerial vehicles (UAVs) has witnessed rapid growth in recent years, driven by technological developments in the sector. These advances have expanded the application of UAVs across numerous industries including construction \cite{GUPTA2023}, agriculture \cite{Kim2019}, infrastructure inspection, monitoring \cite{Savva2021, LEKIDIS20221394}, cinematography \cite{Mademlis2019}, and more. With the rapid increase in UAV operations, especially for complex missions like Beyond Visual Line of Sight (BVLOS) \cite{Sorbelli2024} flights and urban deployments, the need for a structured airspace management system becomes critical. U-space was introduced in the regulation (EU) 2021/664 \cite{eu2021664} and has been developed to be a specialized, automated traffic management framework for UAVs. It integrates digital services such as network identification, geofencing, flight authorization, and conformance monitoring to ensure safe UAV operations. 

Recent research and real-world trials further highlight the impact of U-space. Studies show that U-space enables highly automated and parallel BVLOS operations, essential for scaling large fleets and supporting emerging commercial drone services \cite{Dobrev2024}. The U-space and Unmanned Traffic Management (UTM) frameworks have also been shown to unlock new opportunities for complex missions, such as UAV-based medical transport, by enabling reliable and regulated operation in challenging environments \cite{kotlinski2022u}. Finally, improvements such as contingency management services are essential for volume safety, ensuring robust mitigation of risks and emergencies during routine UAV traffic \cite{munoz2024u}.

By leveraging U-space services, the integration of autonomous UAVs can unlock new operational opportunities and enhance the overall efficiency of airspace utilization. Thanks to on-board autonomy, autonomous UAVs are capable of executing flight trajectories with a higher degree of predictability compared to piloted UAVs. Such predictability enables more accurate forecasting of future positions, thereby supporting a more optimal trade-off between operational efficiency and the safe management of airspace.
Despite this potential, current flight authorization schemes are largely agnostic to the degree of UAV autonomy and typically rely on conservative safety buffers that create overly large reserved volumes. While these buffers guarantee safety, they significantly reduce airspace capacity and limit the number of UAVs that can operate simultaneously. Significant progress has been made in developing U-space infrastructure, yet key challenges persist in optimizing 4D volume reservations for dense urban operations. As urban UAV traffic continues to grow, addressing these issues is essential to improve airspace efficiency, ensure fair access among operators, and maintain safe and effective traffic management.

This is precisely the focus of this work, which aims to determine the minimum 4D operational volumes required for a UAV mission. Specifically, this work's contributions are the following:
\begin{itemize}
\item Development of a 4D volume reservation algorithm that leverages UAV autonomy to optimize flight authorization services within the U-space framework.
\item Demonstration of enhanced airspace utilization and scalability by reserving only the necessary operational volume, enabling simultaneous high-density UAV operations in urban environments.
\item Validation of the proposed method through simulations reflecting realistic flight scenarios, proving its effectiveness and practical potential for operational deployment.
\end{itemize}




The remainder of the paper is organized as follows: Section \ref{related} presents related work. Section \ref{formulation} provides an overview of the proposed solution, while Section \ref{location prediction} discusses the location prediction method. Section \ref{approach} introduces the approach for creating minimum volume flight authorization requests. Section \ref{performance evaluation} showcases the performance evaluation of the system. Finally, Section \ref{conclusion} concludes the paper and discusses possible directions for future research.
\section{Related Work}\label{related}

The concept of U-space represents a coordinated framework of services and infrastructure designed to enable the safe integration of UAVs into shared airspace through the automation of flight operations. Early research has addressed foundational aspects such as regulatory development, flight authorization, conflict management, and communication feasibility, all of which are essential for enabling BVLOS and high-density urban UAV operations \cite{Jepsen2024}. Despite this progress, several obstacles remain. Current flight authorization processes often suffer from delays and lack efficient mechanisms for managing dense and dynamic traffic situations, raising concerns about scalability as UAV activity continues to increase \cite{carraminanascalable}. Although initial U-space services such as registration and geofencing are already deployable, significant challenges persist for real-time operations, including accurate tracking, communication reliability, and situational awareness \cite{Rognin2020}. Moreover, performance differences among U-space Service Providers (USSPs) directly affect efficiency and fairness, emphasizing the need for standardized and transparent processes \cite{munoz2023impact}. Addressing these limitations is critical to achieving a seamless and sustainable integration of high-density UAV operations within the European airspace.

Moreover, a comprehensive U-space Concept of Operations (CONOPS) \cite{COROS-XUAM} has been published, defining the services, infrastructure, and responsibilities required for safe UAV traffic management across both controlled and uncontrolled airspace. Building on this foundation, recent research has proposed operational strategies to support highly automated UAV operations. For example, \cite{gonzalez2025impact} assesses the role of strategic planning in shared U-space volumes, while \cite{Sorbelli2023} introduces a multi-layer framework that incorporates obstacles, no-fly zones, connectivity, and ground risk into BVLOS mission planning. Other efforts, such as \cite{Dobrev2024}, demonstrate parallel BVLOS operations for agricultural monitoring, highlighting the potential of automation in mission execution and traffic scalability. Complementary efforts have focused directly on flight authorization processes. The EuroDrone UTM testbed \cite{Lappas2020} validates interoperability across U-space Levels 1–3, identifying regulatory gaps and technical requirements that affect authorization workflows in urban environments. In parallel, \cite{carraminanascalable} proposes a scalable authorization workflow aimed at ensuring fairness and capacity management, introducing a multi-queue, priority-based service within the Common Information Service Provider (CISP). Their results demonstrate that such mechanisms can mitigate unfairness and prevent monopolization of airspace resources. 

While recent contributions have advanced automation workflows and multi-constraint mission planning, they typically assume conventional flight authorization procedures. In particular, volume reservations are often handled in a static or rule-based manner, resulting in conservative and oversized allocations in space and time. Although such approaches guarantee safety, they lead to inefficient airspace usage, especially in dense urban environments. Initial results in \cite{munoz2023impact} demonstrate this effect: static-rule but high-resolution 4D reservation algorithms nearly double the mission acceptance rate compared to lower-resolution reservations. This underscores the importance of efficient and precise volume reservations for enabling high-density UAV operations and maximizing airspace utilization. To address this limitation, we introduce a novel 4D volume reservation algorithm within the U-space framework. Our approach leverages probabilistic trajectory predictions of autonomous UAVs to dynamically compute the minimum operational volumes required for each mission, ensuring strict containment while substantially reducing reserved airspace. Unlike prior work, this method explicitly exploits UAV autonomy to improve the efficiency of flight authorization, thereby enhancing scalability and capacity in future high-density operations.

\section{Solution Overview}\label{formulation}
By combining enhanced motion predictability with advanced control flexibility, autonomous UAVs provide the foundation for more efficient airspace management. Building on these capabilities, this work introduces a novel framework that generate optimized flight authorizations. The method enables operators to request the \emph{minimum possible 4D volumes}, ensuring with high probability that the UAV remains within the reserved airspace. By reserving only the strictly necessary volumes, the approach contributes to increased airspace capacity and facilitates the simultaneous accommodation of multiple UAV operations.

\begin{figure}[!ht]
\centering
\includegraphics[width=0.9\linewidth]{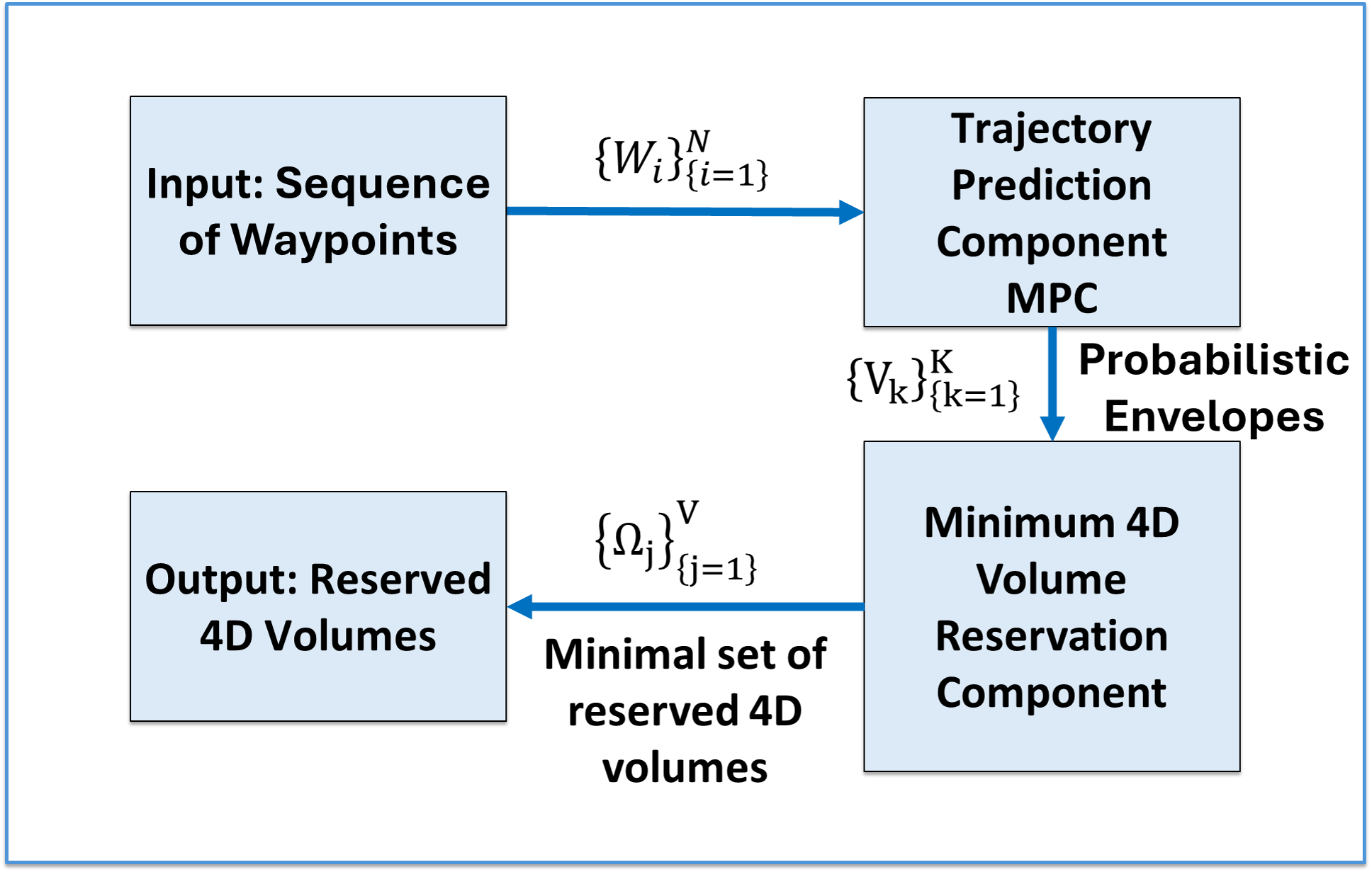}
\caption{System Overview}\label{fig:system}
\end{figure}

\subsection{The 4D Reservation Algorithm}
The \emph{4D volume reservation algorithm} presented in this work is organized into two interconnected components, whose high-level descriptions are provided in the following and whose representation is shown in Fig. \ref{fig:system}.

\subsubsection{The Trajectory Prediction Component}
The first block operates in discrete time slots and receives as input a sequence of desired waypoints defining the intended path of the autonomous UAV, 
\(\mathcal{W} = \{\mathbf{w}_i\}_{i=1}^N\), 
where each waypoint is represented by the vector

\begin{equation}
\mathbf{w}_i = \begin{bmatrix} \mathbf{x}_i^w \\ \dot{\mathbf{x}}_i^w \end{bmatrix} \in \mathbb{R}^6,
\end{equation}
with \(\mathbf{x}_i^w \in \mathbb{R}^3\) denoting the spatial position of the \(i\)-th waypoint and 
\(\dot{\mathbf{x}}_i^w \in \mathbb{R}^3\) the associated desired velocity vector.  
Explicitly, \(\mathbf{x}_i^w = (x_i^w, y_i^w, z_i^w)^\top\) and \(\dot{\mathbf{x}}_i^w = (\dot{x}_i^w, \dot{y}_i^w, \dot{z}_i^w)^\top\).

Given a sequence of desired waypoints, Model Predictive Control (MPC) is employed \emph{offline} to generate feasible state trajectories for the autonomous UAVs. These trajectories are computed under the assumption that the UAV dynamics evolve according to the following model

\begin{equation}
\mathbf{x}_{k+1} = f(\mathbf{x}_k, \mathbf{u}_k,\mathbf{n}_k),
\end{equation}
where the state vector is defined as
\begin{equation}
\mathbf{x}_k = \begin{bmatrix} \mathbf{p}_k \\ \dot{\mathbf{p}}_k \end{bmatrix} \in \mathbb{R}^6,
\end{equation}
with \(\mathbf{p}_k = (x_k, y_k, z_k)^\top\) denoting the UAV position at time step \(k\), 
\(\dot{\mathbf{p}}_k = (\dot{x}_k, \dot{y}_k, \dot{z}_k)^\top\) the velocity, 
and \(\mathbf{n}_k\) a stochastic disturbance term accounting for external factors, e.g., wind effects  and model inaccuracies.  
The control input is denoted by \(\mathbf{u}_k\), and \(f(\cdot)\) is the discrete-time motion model. 

The resulting trajectories are denoted by \(\{\mathbf{x}_k\}_{k=0}^{K}\), with associated control inputs \(\{\mathbf{u}_k\}_{k=0}^{K-1}\), where \(K\) represents the prediction horizon corresponding to the UAV’s mission duration from take-off to landing. The MPC-generated control sequence is obtained by minimizing the traversal time between waypoints.

Explicitly accounting for stochastic disturbances \(\mathbf{n}_k\), the output of this block is a time-indexed sequence of volumes
\(\mathcal{V}_k \subset \mathbb{R}^3\), each representing a high-probability region that contains the UAV’s location at the corresponding time step, i.e.,
\begin{equation}
\mathbb{P}\!\left( \mathbf{p}_k \in \mathcal{V}_k \right) \geq 0.95.
\end{equation}

These volumes represent the spatial envelopes that must be respected by the subsequent reservation block.

\subsubsection{The Minimum 4D Volume Authorization Request Component}
The second block converts the probabilistic envelopes \(\{\mathcal{V}_k\}_{k=0}^{K}\) into a minimum set of reserved 4D volumes 
\(\{\Omega_j\}_{j=1}^{V}\), 
with each \(\Omega_j \subset \mathbb{R}^3 \times \mathbb{R}\) represented as a rectangle in the \((x,y)\)-plane and extending vertically to the U-space ceiling. These volumes define the UAV's reserved airspace. Each volume \(\Omega_j\) is associated with a start time \(s_j\) and an end time \(e_j\), so that it is active for the interval \([s_j,e_j]\). Although not explicitly detailed in this work, it is assumed that an additional operational buffer, as specified by U-space regulations, is applied to each volume when submitting the Flight Authorization Request.

To ensure robustness, each reserved volume must last at least \(k^v\) seconds, and consecutive volumes must overlap in time by at least \(k^o\) seconds, where \(k^v, k^o \in \mathbb{R}^+\) are design parameters defined during U-space designation.

The high-level optimization problem underlying the \textit{minimum-volume flight authorization request} can be summarized as follows
\begin{align}
\min_{\{\Omega_j,s_j,e_j\}_{j=1}^{V}} \quad 
    & \sum_{j=1}^{V} \Omega_j
    \label{eq:objective} \\[0.5em]
\text{s.t.} \quad 
    & \mathcal{V}_k \subseteq \bigcup_{j=1}^{V} \Omega_j, 
    && \forall k,
    \label{eq:coverage} \\
    & e_j - s_j \geq k^v, 
    && \forall j,
    \label{eq:duration} \\
    & e_j - s_{j+1} \geq k^o, 
    && \forall j.
    \label{eq:separation}
\end{align}

\section{Location Prediction for Unmanned Autonomous Vehicles}\label{location prediction}
The first component of the proposed 4D volume reservation solution is responsible for predicting the evolution of the autonomous UAV along the intended sequence of waypoints. To this end, a discrete-time motion model is combined with a MPC strategy that ensures convergence of the UAV state towards the designated waypoints. The resulting state distributions are subsequently used to estimate the volumes \(\mathcal{V}_k\) representing the regions of the airspace in which the UAV is expected to be located.

\subsection{UAV Motion Model}
In this work, a linear stochastic kinematic model is assumed for the UAV, following the formulation presented in \cite{Vitale2022}. Using the state notation introduced in Section \ref{formulation}, the UAV dynamics evolve over one time step as  
\begin{equation}
\mathbf{x}_{k+1} = A \mathbf{x}_k + B \mathbf{u}_k + \mathbf{n}_k,
\label{eq:onestep}
\end{equation}
where $\mathbf{u}_k \in \mathbb{R}^3$ is the control input, represented as the force vector applied to the UAV along the three spatial axes, and $\mathbf{n}_k \in \mathbb{R}^3 \sim \mathcal{N}(0,Q)$ is an i.i.d. Gaussian disturbance with covariance matrix $Q$. The system matrices are given by
\[
A = 
\begin{bmatrix}
I_3 & \Delta T ~I_3 \\
\textit{0}_3 & (1-\eta) ~I_3
\end{bmatrix},
\quad 
B = 
\begin{bmatrix}
\textit{0}_3 \\
\frac{\Delta T}{m}~ I_3
\end{bmatrix},
\]
where $I_3$ and $\textit{0}_3$ are the $3 \times 3$ identity and zero matrices, respectively; $\eta$ is the air resistance coefficient, $m$ denotes the UAV mass, and $\Delta T$ indicates the sampling interval. 

By iterating \eqref{eq:onestep}, the state evolution over \(k\) steps, given the initial UAV state \(\mathbf{x}_0\) and the control sequence \(\{\mathbf{u}_0,\dots,\mathbf{u}_{k-1}\}\), can be expressed as
\begin{equation}
\mathbf{x}_k = A^k \mathbf{x}_0 + \sum_{i=0}^{k-1} A^{k-1-i} B \mathbf{u}_i + \sum_{i=0}^{k-1} A^{k-1-i} \mathbf{n}_i.
\label{eq:state_evolution}
\end{equation}

Since the disturbance terms \(\mathbf{n}_i\) are assumed independent and Gaussian, it follows that the distribution of \(\mathbf{x}_k\) is also Gaussian,
\begin{equation}
\mathbf{x}_k \sim \mathcal{N}(\boldsymbol{\mu}_k, \Sigma_k),
\label{eq:state_distribution}
\end{equation}
\vspace{-0.05in}
\noindent with mean
\vspace{-0.05in}
\begin{equation}
\boldsymbol{\mu}_k = A^k \mathbf{x}_0 + \sum_{i=0}^{k-1} A^{k-1-i} B \mathbf{u}_i,
\label{eq:mean}
\end{equation}
\vspace{-0.05in}
\noindent and covariance
\vspace{-0.05in}
\begin{equation}
\Sigma_k = \sum_{i=0}^{k-1} A^{k-1-i} Q (A^{k-1-i})^\top.
\label{eq:covariance}
\end{equation}

The mean trajectory \(\boldsymbol{\mu}_k\) characterizes the expected evolution of the UAV state under the applied control inputs. More specifically, the position and velocity components are given by \(\boldsymbol{\mu}^p_k\) and \(\boldsymbol{\mu}^v_k\), respectively. The associated covariance matrix \(\Sigma_k\) quantifies the propagation of uncertainty over time resulting from process disturbances. The positional component of the distribution, obtained from \(\boldsymbol{\mu}^p_k\) and the corresponding block of \(\Sigma_k\), serve as the basis for the construction of probabilistic reservation volumes in the subsequent sections.

\subsection{Waypoint-Matching Path Planning}
Given the initial known UAV state $\mathbf{x}_0$ at waypoint $i-1$ and the target waypoint $\mathbf{w}_i$, MPC is used to compute the control inputs $\{\mathbf{u}_k\}_{k=0}^{K_i-1}$ that minimize the deviation of the predicted UAV trajectory from the target over a sufficiently long planning horizon of length $K_i$. Deviations are measured in both position and velocity, and an exponentially increasing weight $\rho^k$ ($\rho>1$) is applied to encourage rapid convergence towards the waypoint. The resulting controls $\{\mathbf{u}_k\}_{k=0}^{K_i-1}$ determine the expected UAV trajectory between waypoints $i-1$ and $i$. Specifically, the following MPC strategy is employed:
\begin{align}
\min_{\{\mathbf{u}_k\}_{k=0}^{K_i-1}} \quad 
    & \sum_{k=0}^{K_i} \rho^k \, \| \boldsymbol{\mu}_k - \mathbf{w}_i \|_2^2,
    \label{eq:mft_mpc_objective} \\[0.5em]
\text{s.t.} \quad 
    & \boldsymbol{\mu}_k = A^k \mathbf{x}_0 + \sum_{i=0}^{k-1} A^{k-1-i} B \mathbf{u}_i, 
    &\quad & \forall k, 
    \label{eq:dyn_constraint} \\
    & \| \mathbf{u}_k \|_\infty \leq u_\text{max}, 
    &\quad & \forall k, 
    \label{eq:max_control} \\
    & \| \mathbf{u}_{k+1} - u_k \|_\infty \leq \Delta u_\text{max}, 
    &\quad & \forall k, 
    \label{eq:max_delta_control} \\
    & \|\boldsymbol{\mu}^v_k \|_\infty \leq v_\text{max}, 
    &\quad & \forall k, 
    \label{eq:max_velocity}
\end{align}
where constraints~\eqref{eq:max_control} and~\eqref{eq:max_delta_control} limit the control input magnitude and its variation between consecutive steps, and where constraint~\eqref{eq:max_velocity} enforces a maximum velocity bound.  

The parameter $K_i$ denotes the discrete time horizon allocated for the UAV to reach waypoint $\mathbf{w}_i$. Within this horizon, the waypoint is considered reached at time step $K_{w_i}$, corresponding to the instant at which the predicted UAV state is closest to $\mathbf{w}_i$, and any remaining control inputs are disregarded. This procedure is applied to all waypoints. A feasible solution ensures that the UAV trajectory closely approaches each waypoint while satisfying dynamic and operational constraints, with exponential weighting promoting rapid convergence towards the target.

\subsection{Location Prediction for Unmanned Autonomous Vehicles}
\label{sec:location_prediction}
Given the MPC strategy presented earlier, the control inputs $\{\mathbf{u}_k\}_{k=0}^{K_{w_i}-1}$ applied by the autonomous UAV to track the intended waypoints are determined, yielding the predicted trajectory. As stated in \eqref{eq:state_distribution}, the state distribution at each time step remains Gaussian, with $\boldsymbol{\mu}_k$ and $\Sigma_k$ computed as in \eqref{eq:mean} and \eqref{eq:covariance}, now fully specified, since the applied controls $\{\mathbf{u}_k\}_{k=0}^{K_{w_i}-1}$ are known. The marginal distribution of the UAV position is therefore
\[
\mathbf{p}_k \sim \mathcal{N}(\boldsymbol{\mu}^p_k, \Sigma^p_k),
\]
where $\boldsymbol{\mu}^p_k$ and $\Sigma^p_k$ denote the mean and covariance of the spatial component of $\mathbf{x}_k$.

In principle, probabilistic envelopes $\{\mathcal{V}_k\}_{k=0}^{K}$ can be obtained by integrating the probability density of $\mathbf{p}_k$, resulting in ellipsoids centered at $\boldsymbol{\mu}^p_k$ \cite{ribeiro2004gaussian}. However, within the U-space framework, reserved volumes $\Omega_j$ are assumed to extend vertically to the ceiling. Consequently, the $z$-dimension does not affect the optimization and the formulation can be restricted to the $(x,y)$-plane. Let $\boldsymbol{\mu}^{xy}_k$ and $\Sigma^{xy}_k$ denote the mean and covariance of the UAV position projected onto the horizontal plane. The ellipse that characterizes the region containing the UAV with a specified confidence level, $95\%$ in the following, can be constructed from the eigenvalues $\lambda^{xy}_{k,1}$ and $\lambda^{xy}_{k,2}$ of $\Sigma^{xy}_k$. Specifically, the eigenvalues determine the squared semi-axes of the confidence ellipse \cite{ribeiro2004gaussian}:

\begin{align}
a^{xy}_{k,1} &= \sqrt{\chi^2_{2,0.95}\,\lambda^{xy}_{k,1}}, \nonumber \\
a^{xy}_{k,2} &= \sqrt{\chi^2_{2,0.95}\,\lambda^{xy}_{k,2}}, 
\label{eq:ellipse}
\end{align}
where $\chi^2_{2,0.95}$ is the $95$th percentile of the chi-squared distribution with two degrees of freedom.

For tractability, each ellipse is conservatively approximated by a circle of radius
\begin{equation}
r_k = \max\{a^{xy}_{k,1},\,a^{xy}_{k,2}\},
\label{eq:circle}
\end{equation}
centered at $\boldsymbol{\mu}^{xy}_k$. This worst-case approximation guarantees that the true confidence ellipse is fully contained within the circular footprint, while substantially simplifying the subsequent optimization.

\section{An Approach for Minimum-Volume Flight Authorization Requests}\label{approach}
Building on the predicted UAV footprints $\{(\boldsymbol{\mu}^{xy}_k,r_k)\}_{k=0}^{K}$ derived in Sec.~\ref{location prediction}, this section presents a framework for generating \textit{minimum-volume flight authorization requests}. The approach approximates the probabilistic spatial envelopes of the UAV trajectory with a set of compact 4D volumes, ensuring safe containment while minimizing the total reserved airspace.

\subsection{MILP Formulation for Minimum-Volume Flight Authorization Requests}
The task of minimizing the total reserved 4D space-time is formulated as a mixed-integer linear program (MILP). Each predicted UAV footprint at time step $k$ must be fully contained within at least one reserved 4D volume $\Omega_j \subset \mathbb{R}^3 \times \mathbb{R}$. As previously noted, each \(\Omega_j\) is represented as a rectangle in the \((x,y)\)-plane, extending vertically to the U-space ceiling, and is associated with an activation time interval. Formally, each 4D volume \(\Omega_j\) is parameterized by its spatial bounds \((x_\text{min}^j, x_\text{max}^j, y_\text{min}^j, y_\text{max}^j)\) and its temporal interval \([s_j,e_j]\). To improve packing efficiency, a discrete set of orientations is considered. In this work, two orientations ($0^\circ$ and $45^\circ$) relative to the Cartesian axes of the airspace are incorporated, although the approach can be generalized to additional angles. An auxiliary binary variable $\delta_j$ specifies whether $\Omega_j$ is rotated by $45^\circ$ ($\delta_j=1$) or remains axis-aligned ($\delta_j=0$). 

Based on these definitions, the problem of minimizing the total reserved 4D airspace, while ensuring full coverage of the UAV trajectory under temporal and orientation constraints, is formulated as the following MILP:

\begin{subequations}
\begin{alignat}{3}
&\rlap{\textbf{Problem } \texttt{Min Volume Reservation}:} & & & \nonumber\\[-1mm]
& \min \sum_k \sum_j z_{kj} A_j & & \label{eq:opta}\\[1mm]
& \text{decision variables: } && \nonumber\\
& z_{kj} \in \{0,1\}, ~\delta_j \in \{0,1\}, && \forall k,j \nonumber\\
& x_\text{min}^j, x_\text{max}^j, y_\text{min}^j, y_\text{max}^j, && \forall j \label{eq:optb}\\
&\text{subject to:} & & & \nonumber\\[-1mm]
& b_{kj} \le z_{kj}, \quad b_{kj} + \delta_j \le 1, \quad b_{kj} \ge z_{kj} - \delta_j &\quad & \forall k,j \label{eq:optc}\\[-1mm]
& x_\text{min}^j \le \mu^{x}_k - r_k + M (1-b_{kj}) &\quad & \forall k,j  \label{eq:optd}\\[-1mm]
& x_\text{max}^j \ge \mu^{x}_k + r_k - M (1-b_{kj}) &\quad & \forall k,j \label{eq:opte} \\[-1mm]
& y_\text{min}^j \le \mu^{y}_k - r_k + M (1-b_{kj}) &\quad & \forall k,j \label{eq:optf}\\[-1mm]
& y_\text{max}^j \ge \mu^{y}_k + r_k - M (1-b_{kj}) &\quad & \forall k,j \label{eq:optg}\\[-1mm]
& x_\text{min}^j \le \mu^x_{k,\delta} - r_k + M (1-(z_{kj}-b_{kj})) &\quad & \forall k,j \label{eq:opth}\\[-1mm]
& x_\text{max}^j \ge \mu^x_{k,\delta} + r_k - M (1-(z_{kj}-b_{kj})) &\quad & \forall k,j \label{eq:opti}\\[-1mm]
& y_\text{min}^j \le \mu^y_{k,\delta} - r_k + M (1-(z_{kj}-b_{kj})) &\quad & \forall k,j \label{eq:optj}\\[-1mm]
& y_\text{max}^j \ge \mu^y_{k,\delta} + r_k - M (1-(z_{kj}-b_{kj})) &\quad & \forall k,j \label{eq:optk}\\[-1mm]
& A_j = w_j h_j &\quad & \forall j \label{eq:optl}\\[-1mm]
& w_j \ge x_\text{max}^j - x_\text{min}^j &\quad & \forall j \label{eq:optm}\\[-1mm]
& h_j \ge y_\text{max}^j - y_\text{min}^j  &\quad & \forall j \label{eq:optn}\\[-1mm]
& \sum_j z_{kj} \ge 1 &\quad & \forall k \label{eq:opto}\\[-1mm]
& \sum_k z_{kj} \ge k^v &\quad & \forall j \label{eq:optp}\\[-1mm]
& s_j \le k z_{kj} + M (1-z_{kj}), \quad e_j \ge k z_{kj} &\quad & \forall k,j \label{eq:optq}\\[-1mm]
& \sum_k z_{kj} = e_j - s_j + 1 &\quad & \forall j \label{eq:optr}\\[-1mm]
& e_j \ge s_j, \quad s_{j+1} \ge s_j, \quad e_{j+1} \ge e_j &\quad & \forall j \label{eq:opts}\\[-1mm]
& e_j - s_{j+1} \ge k^o &\quad & \forall j\label{eq:optt}
\end{alignat}
\end{subequations}

In the \texttt{Min Volume Reservation} optimization, each assignment variable \(z_{kj} \in \{0,1\}\) indicates whether the horizontal footprint of the UAV at time step \(k\) is covered by the 4D volume \(\Omega_j\). By construction, \(z_{kj}=1\) if footprint \(k\) is contained within \(\Omega_j\), and \(z_{kj}=0\) otherwise. Hence, the objective function \eqref{eq:opta} minimizes the total reserved airspace by summing over all assignments. Specifically, the horizontal footprint associated with \(\Omega_j\) is counted once for each envelope \(k\) it covers, so that its contribution to the objective is proportional to the number of time steps during which it is active. Indeed, multiplying \(z_{kj}\) by the area \(A_j\) of the horizontal footprint of \(\Omega_j\) captures this space-time cost; the horizontal area is weighted by the temporal duration implied by the sum over \(k\), effectively accounting for the total 4D volume reserved by \(\Omega_j\).

The coverage and rotation logic is encoded via the auxiliary variable \(b_{kj}\) in \eqref{eq:optc}. By construction, \(b_{kj} = 1\) when the axis-aligned constraints are active and \(b_{kj} = 0\) when the rotated constraints are active. Indeed, the three inequalities in \eqref{eq:optc}, together with the large constant \(M\) in the Big-M formulation of \eqref{eq:optd}--\eqref{eq:optk}, ensure that the appropriate set of coverage constraints is enforced. Specifically, when \(z_{kj} = 1\), if \(\delta_j = 0\), the third inequality in \eqref{eq:optc} forces \(b_{kj} = 1\), activating the axis-aligned constraints \eqref{eq:optd}--\eqref{eq:optg}; conversely, if \(\delta_j = 1\), the second inequality in \eqref{eq:optc} forces \(b_{kj} = 0\), activating the rotated constraints \eqref{eq:opth}--\eqref{eq:optk}. The Big-M terms relax the inactive constraints, ensuring they do not restrict the solution and guaranteeing that exactly one set of constraints is active for each footprint assignment, consistent with the chosen orientation of \(\Omega_j\). When \(z_{kj} = 0\), the UAV footprint \(k\) is not assigned to volume \(\Omega_j\). In this case, \(b_{kj} = 0\) by the first inequality in \eqref{eq:optc}, and the corresponding coverage constraints are inactive, rendering them irrelevant to the feasibility of the optimization.

The axis-aligned constraints \eqref{eq:optd}--\eqref{eq:optg} enforce that the predicted horizontal footprint of the UAV at time \(k\), represented by \((\mu^x_k, \mu^y_k)\) with radius \(r_k\), is fully contained within the horizontal bounds \([x_\text{min}^j, x_\text{max}^j] \times [y_\text{min}^j, y_\text{max}^j]\) of volume \(\Omega_j\). The rotated constraints \eqref{eq:opth}--\eqref{eq:optk} are obtained by applying a planar rotation matrix \(R_{\theta}\) to the footprint coordinates
\[
\begin{bmatrix} \mu^x_{k,\delta} \\ \mu^y_{k,\delta} \end{bmatrix} = 
R_{\theta} \begin{bmatrix} \mu^x_k \\ \mu^y_k \end{bmatrix}, \quad
R_{\theta} = \begin{bmatrix} \cos \theta & -\sin \theta \\[1mm] \sin \theta & \cos \theta \end{bmatrix}, \quad \theta = 45^\circ.
\]
The resulting coordinates \((\mu^x_{k,\delta}, \mu^y_{k,\delta})\) represent the horizontal footprint of the UAV in the rotated reference frame of \(\Omega_j\). The rotated constraints then ensure that this rotated footprint is fully contained within the corresponding horizontal bounds \([x_\text{min}^j, x_\text{max}^j] \times [y_\text{min}^j, y_\text{max}^j]\) of the volume.

Constraint \eqref{eq:optl} defines the area of each rectangle \(A_j\) as the product of its width \(w_j\) and height \(h_j\). The additional constraints \eqref{eq:optm}--\eqref{eq:optn} enforce that \(w_j\) and \(h_j\) are at least as large as the horizontal extents of the rectangle. Introducing \(w_j\) and \(h_j\) as explicit variables simplifies the optimization problem by avoiding direct multiplication of decision variables in the objective function and constraints, while still guaranteeing that the calculated area \(A_j\) fully contains the assigned footprint in any orientation.

Constraint \eqref{eq:opto} ensures that every footprint \(k\) is covered by at least one volume. Constraint \eqref{eq:optp} enforces a minimum active duration \(k^v\) for each volume. The start time \(s_j\) and end time \(e_j\) of each volume, defined by \eqref{eq:optq}--\eqref{eq:optr}, correspond respectively to the first and last time steps at which the volume \(\Omega_j\) is active. In particular, \(s_j\) becomes equal to the minimum \(k\) for which the Big-M terms in the coverage constraints \eqref{eq:optd}--\eqref{eq:optk} become active, effectively marking the first activation of the volume, while \(e_j\) is equal to the maximum \(k\) for which the volume is assigned to any footprint. Constraint \eqref{eq:optr} further enforces that the assignments of volume \(\Omega_j\) are consecutive in time, so that there are no gaps between its activations. Temporal consistency and ordering across volumes are imposed by \eqref{eq:opts}, and the minimum temporal overlap \(k^o\) between consecutive volumes is enforced by \eqref{eq:optt}.

Finally, the total number of reserved volumes, $V$, serves as a tunable parameter rather than a decision variable within the MILP. While a higher $V$ theoretically allows for a tighter approximation of the flight path, the aforementioned safety constraints regarding the minimum volume duration $k^v$ and the mandatory temporal overlap $k^o$ impose a limit on airspace fragmentation. Consequently, increasing $V$ indefinitely yields diminishing returns, and in practice, a small number of volumes suffices for typical urban trajectories. To identify the optimal count, a sequential sensitivity analysis is performed: the \texttt{Min Volume Reservation} optimization is solved iteratively with increasing values of $V$ until the objective function converges, indicating that maximum packing efficiency has been reached.

Overall, the presented framework provides a rigorous mechanism for generating minimum-volume flight authorization requests. By jointly optimizing footprint assignment, volume orientation, and spatial-temporal bounds, it ensures that all predicted UAV positions are fully covered, while maintaining temporal continuity and satisfying minimum-duration requirements. The explicit handling of axis-aligned and rotated footprints, combined with auxiliary variables and Big-M activation, guarantee that each volume is efficiently used in both space and time. The resulting 4D reservation is compact, safe, and minimizes the total reserved volume required for the UAV trajectory.

Clearly, the proposed MILP formulation can involve a large number of variables and constraints, particularly for long trajectories or fine temporal discretizations. However, the optimization is performed \textit{offline}, prior to UAV take-off, so real-time performance is not required. This allows solving the problem in reasonable time using standard MILP solvers, even for complex trajectories.

\section{Performance Evaluation}\label{performance evaluation}
This section evaluates the effectiveness of the proposed \texttt{Min Volume Reservation} framework through simulation experiments. The objectives are to quantify reductions in reserved airspace volume compared to standard baseline strategies and to validate that the probabilistic envelopes derived from UAV trajectory predictions accurately capture the actual paths followed by autonomous UAVs.

\subsection{Simulation Environment}
Two representative UAV flight scenarios are considered:
\begin{enumerate}
    \item Long-Range Trajectory: a straight-line mission connecting a sequence of waypoints at constant speed, representing typical long-distance transit operations.
    \item Circular Monitoring Trajectory: a closed-loop mission along a circular path, where the UAV hovers at each waypoint to perform monitoring tasks.
\end{enumerate}

The main physical parameters, control settings, and reservation-related constraints used throughout the experiments are summarized in Table~\ref{tab:sim_params}.

\begin{table}[t]
\centering
\caption{Simulation Parameters}
\label{tab:sim_params}
\begin{tabular}{lcl}
\toprule
\textbf{Parameter} & \textbf{Symbol / Value} \\
\midrule
Sampling time & $\Delta T = 1 \,\text{s}$ \\
Drag coefficient & $\eta = 0.1$ \\
UAV mass & $m = 10 \,\text{kg}$ \\
Maximum force & $u_\text{max} = 300 \,\text{N}$ \\
Maximum force variation & $\Delta u_\text{max} = 10 \,\text{N}$ \\
Maximum velocity & $v_\text{max} = 14 \,\text{m/s}$ \\
Increasing Exponential Factor & $\rho = 
1.05$ \\
Probability level & $95\%$ \\
Minimum duration & $k^v = 60 \,\text{s}$ \\
Minimum overlap & $k^o = 20 \,\text{s}$ \\
Process noise covariance & $Q = \begin{bmatrix} 0.1I_3 & 0.2I_3 \\ 0.2I_3 & 0.4I_3 \end{bmatrix}$ \\
\bottomrule
\end{tabular}
\end{table}

For the long-range trajectory, the UAV performs a vertical take-off from the origin, climbs to a cruising altitude of $100\,\text{m}$, and traverses seven waypoints at a constant speed of $7\,\text{m/s}$, before executing a vertical landing at $(1500,900,0)$. The mission spans approximately $2\,\text{km}$ and lasts about $3$ minutes and $20$ seconds.

For the circular monitoring trajectory, the UAV similarly ascends vertically to $100\,\text{m}$ and then follows a closed-loop path connecting seven waypoints arranged in a near-circular configuration. At each waypoint, the UAV hovers to perform monitoring tasks before reaccelerating towards the next location, and concludes with a vertical landing at the origin. Although the total path length is slightly below $2\,\text{km}$, the overall mission time is longer (almost $4$ minutes) due to repeated hovering and acceleration phases.

\begin{figure*}[t!]
    \centering
    \begin{subfigure}[b]{0.32\textwidth}
        \centering
        \includegraphics[width=\textwidth]{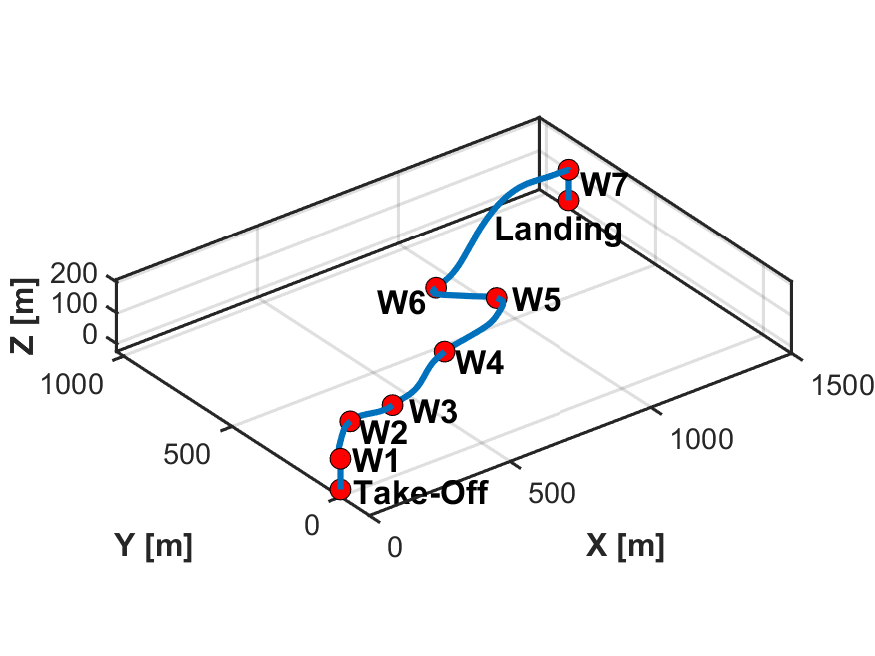}
        \caption{Predicted UAV trajectory.}
        \label{fig:long_predicted}
    \end{subfigure}
    \hfill
    \begin{subfigure}[b]{0.32\textwidth}
        \centering
        \includegraphics[width=\textwidth]{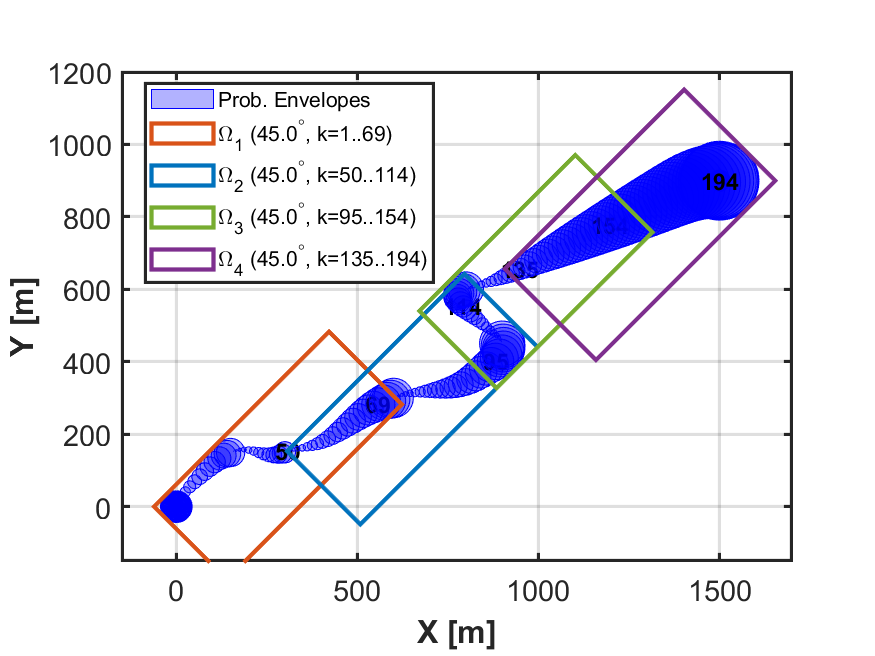}
        \caption{\texttt{Min Volume Reservation} optimization.}
        \label{fig:long_min_volume}
    \end{subfigure}
    \hfill
    \begin{subfigure}[b]{0.32\textwidth}
        \centering
        \includegraphics[width=\textwidth]{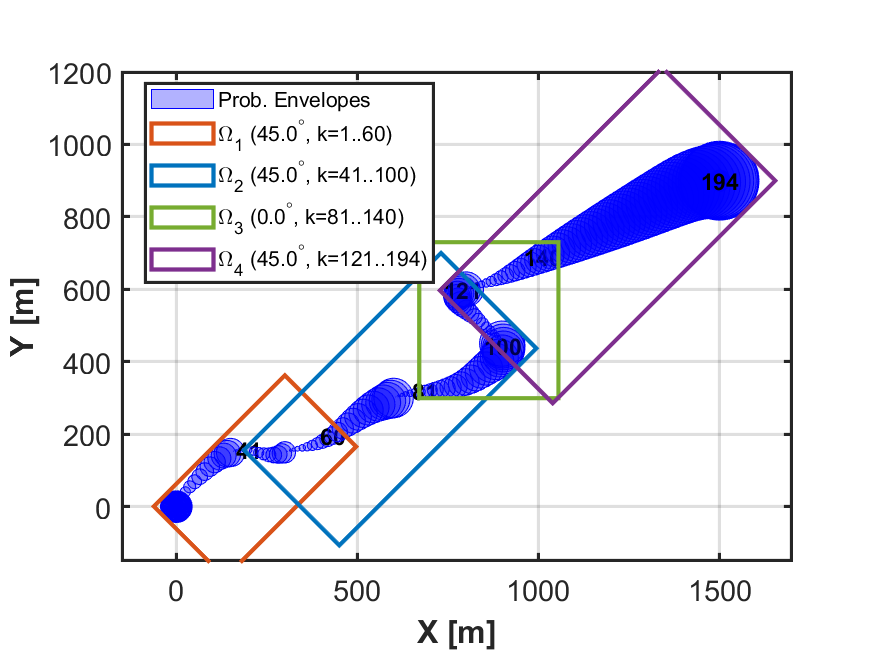}
        \caption{Baseline approach.}
        \label{fig:long_baseline}
    \end{subfigure}

    \caption{Long-Range Trajectory.}
    \label{fig:long_trajectory_comparison}
\end{figure*}

\begin{figure*}[t!]
    \centering
    \begin{subfigure}[b]{0.32\textwidth}
        \centering
        \includegraphics[width=\textwidth]{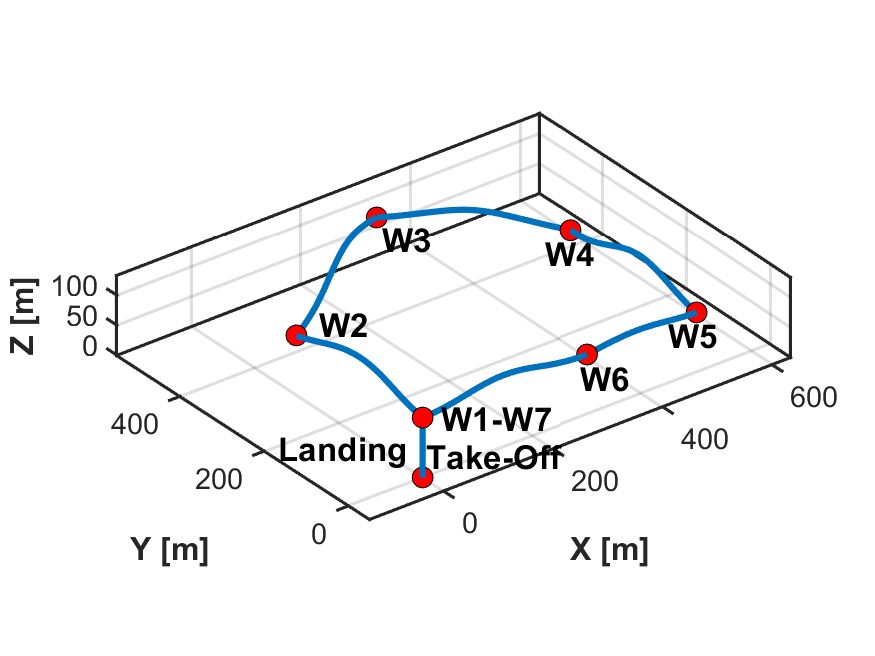}
        \caption{Predicted UAV trajectory.}
        \label{fig:circle_predicted}
    \end{subfigure}
    \hfill
    \begin{subfigure}[b]{0.32\textwidth}
        \centering
        \includegraphics[width=\textwidth]{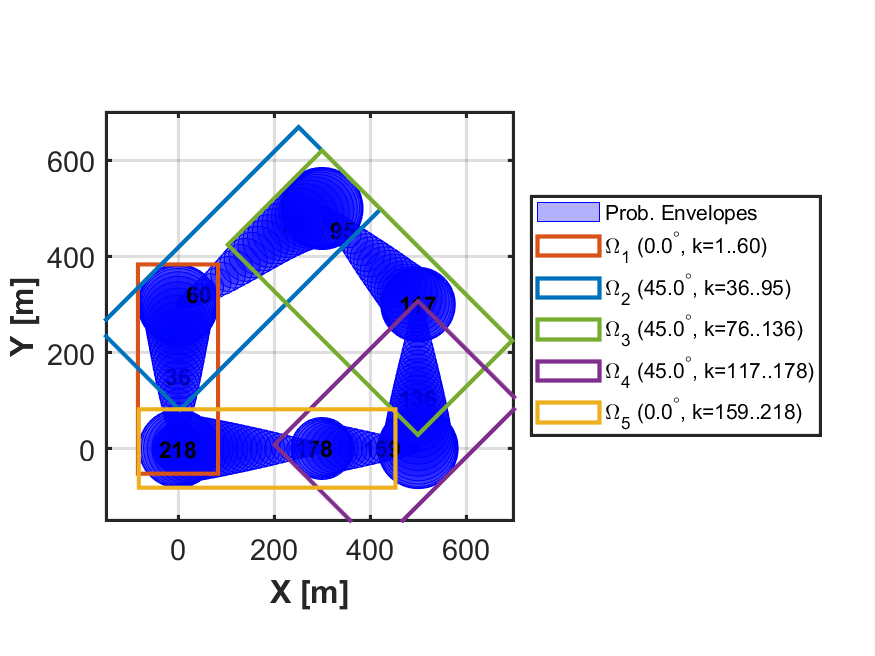}
        \caption{\texttt{Min Volume Reservation} optimization.}
        \label{fig:circle_min_volume}
    \end{subfigure}
    \hfill
    \begin{subfigure}[b]{0.32\textwidth}
        \centering
        \includegraphics[width=\textwidth]{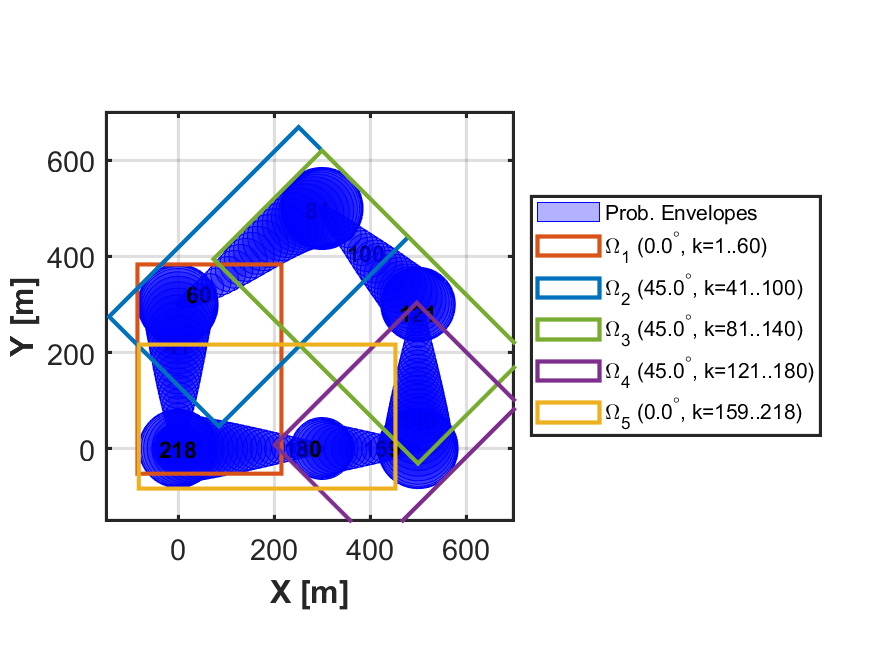}
        \caption{Baseline approach.}
        \label{fig:circle_baseline}
    \end{subfigure}
    \caption{Circular Monitoring Trajectory.}
    \label{fig:circle_trajectory_comparison}
\end{figure*}

\subsection{Performance Assessment and Baseline Comparison}
The effectiveness of the \texttt{Min Volume Reservation} framework is assessed by illustrating the predicted UAV trajectories along with the corresponding 4D reserved volumes for the considered simulation scenarios. 

To provide a performance benchmark, a baseline strategy is introduced in which 4D volumes are generated without solving the \texttt{Min Volume Reservation} optimization problem. Given the expected trajectory $\{\boldsymbol{\mu}^{xy}_k\}_{k=0}^{K}$ of the autonomous UAV, as computed in Section~\ref{location prediction}, the mission is partitioned uniformly in time rather than adapting the 4D volumes to the predicted probabilistic envelopes. Specifically, consecutive temporal segments of length $k^v - k^o$ are defined, ensuring that each reserved volume satisfies both the minimum active duration $k^v$ and the required temporal overlap $k^o$. Furthermore, each volume is modeled with a fixed lateral extent of $300$\,m around the predicted trajectory. While this baseline approach leverages trajectory predictions developed in this work, it remains representative of conventional rule-based reservation practices in U-space and serves as a reference for quantifying the benefits achieved by the proposed optimization framework.

\begin{figure}[!ht]
    \centering
    \includegraphics[width=0.9\linewidth]{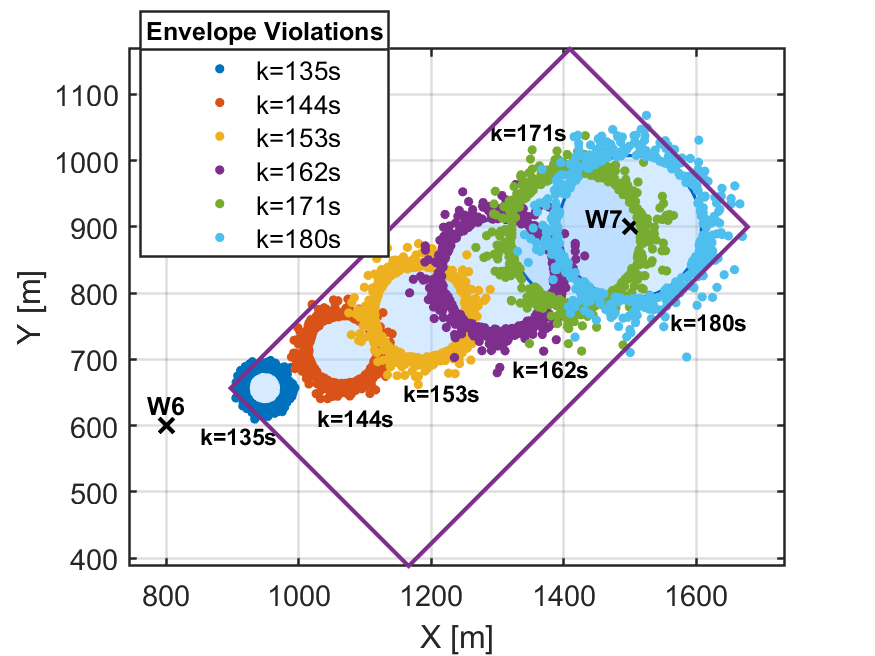}
    \caption{Evaluation of the long-range mission between $W_6$ and $W_7$ against the probabilistic envelopes and reserved 4D volumes.}
    \label{fig:montecarlo}
\end{figure}

Figure~\ref{fig:long_trajectory_comparison} illustrates the results for the long-range mission. Specifically,  Fig.~\ref{fig:long_predicted} depicts the expected UAV trajectory, while Fig.~\ref{fig:long_min_volume} shows the 4D reservation volumes obtained with the proposed \texttt{Min Volume Reservation} framework. Several configurations with different numbers of volumes $\Omega_j$ are tested, and the solution achieving the smallest overall reservation volume is reported.  

The use of probabilistic envelopes of the UAV trajectory enables the optimizer to adapt both the timing and the size of the 4D volumes, leading to substantially more compact authorizations compared to the baseline approach (Fig.~\ref{fig:long_baseline}). In the baseline approach, the mission is evenly partitioned in time and volumes are requested without leveraging trajectory uncertainty, resulting in unnecessarily large reservations. By contrast, the optimized reservations demonstrate that adaptively selecting the start and end times of each volume in accordance with the predicted trajectory significantly reduces the required spatio-temporal authorization. Overall, the requested 4D reservation is reduced by $17.3\%$ relative to the baseline approach, while still guaranteeing complete coverage of the UAV path.

Figure~\ref{fig:circle_trajectory_comparison} presents the results for the circular monitoring mission. The predicted trajectory with hovering phases is shown in Fig.~\ref{fig:circle_predicted}, while the corresponding optimized reservation is depicted in Fig.~\ref{fig:circle_min_volume}. The results demonstrate that the proposed \texttt{Min Volume Reservation} framework adapts seamlessly to mission profiles that include intermittent hovering and repeated accelerations, automatically generating compact authorization requests. In contrast, the baseline approach (Fig.~\ref{fig:circle_baseline}) produces overly conservative volumes that do not exploit the structure of the trajectory. In this scenario as well, the total requested 4D reservation is reduced by $25.1\%$ compared to the baseline approach, while still ensuring full coverage of the UAV path.

Overall, these results highlight the ability of the proposed framework to generate compact and adaptive flight authorizations across diverse mission types, while fully satisfying spatial and temporal constraints. Most notably, in both missions, the reduction in 4D reserved volume is achieved without increasing the number of volumes in the authorization request, but rather by optimally selecting their sizes and activation times.

\subsection{Evaluation of the Approach Robustness}
The robustness of the proposed framework is assessed on the segment between waypoints $W_6$ and $W_7$ of the long-range trajectory. In this scenario, the UAV follows the optimal controls computed in Section~\ref{location prediction}, applied to the motion model that includes the disturbance term $\mathbf{n}_k$. To account for the stochastic effects of this disturbance, a Monte Carlo simulation with $10{,}000$ realizations is performed, generating an ensemble of possible UAV trajectories. Each trajectory is compared against the $95\%$ probabilistic envelopes of the UAV position, which are precomputed as described in Section~\ref{sec:location_prediction}. For clarity of Fig.~\ref{fig:montecarlo}, the envelopes are displayed only at selected time steps rather than continuously.  

As expected, about $5\%$ of the simulated trajectories lie outside the probabilistic envelopes, consistent with their confidence level. It is important to note that this does not mean that the UAV spends $5\%$ of its flight outside the probabilistic envelopes. Rather, at each time step there is a $5\%$ probability of being outside the envelope.  

The reserved 4D volume that covers this segment of the mission is also shown in Fig.~\ref{fig:montecarlo}. Only a small portion of the simulated trajectories (about 0.08\%) extend beyond its boundaries, and those that do remain very close to the edge. This confirms the compact and reliable definition of the reserved volumes achieved by the proposed framework. 

\section{Conclusion}\label{conclusion}
This work introduces a novel optimization framework for generating compact 4D flight authorizations for autonomous UAV missions. The approach leverages trajectory predictions and probabilistic envelopes to adaptively define the spatial and temporal extent of each reserved volume, thereby minimizing the overall reservation while ensuring complete coverage of the UAV path. A baseline strategy is also considered for comparison, and results demonstrated that the proposed method consistently yields significantly smaller and more efficient reservations across different mission types. Crucially, this framework is made possible by the enhanced predictability and reliability of autonomous UAV operations. By exploiting accurate trajectory forecasts, it becomes feasible to request compact authorizations that remain valid and safe, avoiding the overly conservative buffers typically required in conventional rule-based practices.   

Future work will investigate extensions that exploit the versatility of the proposed framework. In particular, ground risk will be incorporated, along with alternative objectives when defining both the UAV path and the corresponding 4D flight authorization request, further broadening the applicability of the approach to diverse operational settings.

\bibliographystyle{IEEEtran}
\bibliography{IEEEabrv,ref}

\end{document}